\documentstyle[12pt,a4,graphicx]{article}

\newcommand{\be}{\begin{equation}}
\newcommand{\ee}{\end{equation}}
\newcommand{\ba}{\begin{eqnarray}}
\newcommand{\ea}{\end{eqnarray}}
\newcommand{\dis}{\displaystyle}

\begin{document}
\begin{titlepage}
\vspace*{-4cm}
\begin{flushright}
CAFPE-40/04\\
IFIC/04-68\\
FTUV/04-1120\\
MPP-2004-146\\
TUM-HEP-568/04\\
UG-FT-170/04\\
\end{flushright}
\vspace{1cm}
\begin{center}

{\large\bf  Extraction of $m_s$ and $|V_{us}|$
from Hadronic Tau Decays
\footnote{
Invited talk given by J.P. 
at `` TAU04, 14-17 September 2004, Nara, Japan''.}}\\
\vspace{1cm}
{\bf Elvira G\'amiz$^{a)}$, 
Matthias Jamin$^{b)}$,
Antonio Pich$^{c)}$,
 Joaquim Prades$^{d)}$ 
and  Felix Schwab$^{b, e)}$ 
}\\[0.5cm]
$^{a)}$ Department of Physics \& Astronomy,
University of Glasgow\\ Glasgow G12 8QQ, United Kingdom.\\[0.5cm]

$^{b)}$ Physik Department, Technische
Universit\"at M\"unchen\\ D-85747 Garching, Germany.\\[0.5cm]

$^{c)}$ Departament de F\'{\i}sica Te\`orica, IFIC, 
Universitat de Val\`encia--CSIC\\ Apt. de Correus 22085, E-46071
Val\`encia, Spain.\\[0.5cm]

$^{d)}$ Centro Andaluz de F\'{\i}sica de las Part\'{\i}culas
Elementales (CAFPE) and Departamento de
 F\'{\i}sica Te\'orica y del Cosmos, Universidad de Granada \\
Campus de Fuente Nueva, E-18002 Granada, Spain.\\[0.5cm]

$^{e)}$ Max-Planck-Institut f\"ur Physik --
Werner-Heisenberg-Institut\\ D-80805 M\"unchen, Germany.\\[0.5cm]
\end{center}
\vfill
\begin{abstract}
\noindent
We review recent work to determine the strange quark mass
$m_s$ as well as  the  proposal to determine $|V_{us}|$ using
hadronic $\tau$ decay data.   The recent update of the
strange spectral function by OPAL and their  moments
of the invariant mass distribution are  employed.
Our results are $|V_{us}|=0.2208\pm0.0034$ and
$m_s(2\,{\rm GeV})=81\pm22$ MeV. Our result is already 
competitive to the standard extraction of $|V_{us}|$ from 
$K_{e3}$ decays and to the new proposals
to determine it. The error on $|V_{us}|$
is dominated by experiment and will be eventually much
improved by the B-factories hadronic $\tau$ data.
 Ultimately, a simultaneous fit of both $m_s$
and $|V_{us}|$ to a set of moments of 
the hadronic $\tau$ decays invariant mass distribution 
will provide one of the most accurate determinations
of these Standard Model parameters. 
\end{abstract}
\vfill
November 2004
\end{titlepage}

\section{Introduction: Theoretical Framework}
 The high precision status reached by 
\be
\label{Rtau}
R_\tau \equiv 
\frac{\Gamma\left[\tau^- \to {\rm hadrons} (\gamma)\right]}
{\Gamma\left[ \tau^- \to e^- \overline{\nu}_e \nu_\tau (\gamma)\right]}
\ee
and related observables
 using the data of the LEP  experiments ALEPH \cite{ALEPH98}
and OPAL \cite{OPAL99}  at CERN
 and the CESR experiment CLEO  \cite{CLEO93} at Cornell
 can be used to determine fundamental QCD parameters \cite{BNP92}.
 Indeed, the analysis of the non-strange inclusive data
 has led to accurate measurements of  
$a_\tau\equiv\alpha_s(M_\tau)/\pi$ which
complement and compete with the current world average
\cite{ALEPH98,OPAL99}. 

The SU(3) breaking induce sizable corrections
 in the semi-inclusive $\tau$-decay  width into
Cabibbo-suppressed modes,  which 
can be analyzed with precision determinations  of
the strange spectral function
\cite{ALEPH99,CLEO03,OPAL04}, 
providing  accurate measurements of
the strange quark mass $m_s(M_\tau)$ 
\cite{CHD97,PP98,CKP98,PP99,KM00,CDGHPP01,GJPPS03,GJPPS04a}.

 More recently, it has been 
pointed out that precision determinations
of the strange spectral function
 can be used to obtain an accurate value
 for the Cabibbo--Kobayashi--Maskawa matrix element
module $|V_{us}|$ \cite{GJPPS03,GJPPS04a,JAM03}. 
 The advantage of this method is that the
experimental uncertainty is expected to be reduced drastically
at the present B-factories: BABAR and BELLE.

 The basic objects one needs to perform the QCD analysis
of (\ref{Rtau}) and related observables
are  Green's two-point functions for vector 
$V^{\mu}_{ij} \equiv \overline{q}_i \gamma^\mu q_j$ and axial-vector
$A^{\mu}_{ij} \equiv \overline{q}_i \gamma^\mu \gamma_5 q_j$ 
color singlets, 
\ba
\Pi^{\mu\nu}_{V,ij}(q)\equiv i \int {\rm d}^4 x \, e^{i q\cdot x}
\langle 0 | T \left([V^\mu_{ij}]^\dagger(x) V^\nu_{ij}(0)\right)
| 0 \rangle \, , \nonumber \\
\Pi^{\mu\nu}_{A,ij}(q)\equiv i \int {\rm d}^4 x \, e^{i q\cdot x}
\langle 0 | T \left([A^\mu_{ij}]^\dagger(x) A^\nu_{ij}(0)\right)
| 0 \rangle \, .
\ea
 The subscripts $i,j$ denote light  quark flavors (up, down
and strange). These correlators admit the Lorentz decompositions
\ba
\Pi^{\mu\nu}_{ij,V/A}(q) &=&
\left(-g_{\mu\nu} q^2 + q^\mu q^\nu \right) \, \Pi^T_{ij,V/A}(q^2)
\nonumber \\
&+&q^\mu q^\nu \, \Pi^L_{ij,V/A}(q^2)\, 
\ea
where the superscripts  in the transverse and longitudinal
components denote the spin $J=1 \, (T)$ and $J=0 \, (L)$
in the hadronic rest frame.

Using the analytic properties of $\Pi^J(s)$, 
one can express $R_\tau$ as a contour
integral in the complex s-plane running counter-clockwise around
the circle $|s|=M_\tau^2$
\ba
R_\tau &\equiv& -i \pi \oint_{|s|=M_\tau^2}
\, \frac{{\rm d} s}{s} \, \left[ 1-\frac{s}{M_\tau^2}\right]^3
\nonumber \\ &\times&
\left\{3 \left[1+\frac{s}{M_\tau^2}\right] \, D^{L+T}(s)
+ 4 \, D^L(s) \right\} \, .
\ea
We have used integration by parts to rewrite $R_\tau$
in terms of the logarithmic derivatives of the relevant
correlators
\ba
D^{L+T}(s)&\equiv& - s \frac{\rm d}{{\rm d} s}
[\Pi^{L+T}(s)] \, ; \nonumber \\ 
D^{L}(s)&\equiv& \frac{s}{M_\tau^2} \frac{\rm d}{{\rm d} s}
[s \Pi^{L}(s)] \, ,
\ea
 which satisfy homogeneous renormalization group
equations, eliminate unwanted renormalization scheme
dependent subtraction constants in $\Pi^L(s)$ and
produce a triple zero in the real axis.
For large enough Euclidean $Q^2$, the correlators $\Pi^{L+T}(Q^2)$
and $\Pi^L(Q^2)$ can be organized in series of dimensional operators
using the operator product expansion (OPE) in QCD.

Moreover, we can decompose $R_\tau$ into
\be
R_\tau \equiv R_{\tau,V} + R_{\tau,A} + R_{\tau,S} 
\ee
according to the quark content
\ba
\Pi^J(s)&\equiv& |V_{ud}|^2 
\left\{ \Pi^J_{V,ud}(s) + \Pi^J_{A,ud}(s) \right\} 
 \nonumber \\ &+& 
|V_{us}|^2 \left\{ \Pi^J_{V,us}(s) + \Pi^J_{A,us}(s) \right\} \, . 
\ea

Additional information can be obtained from the measured
invariant mass distribution of the final state hadrons. 
The corresponding moments 
\be
\label{OPE}
R_\tau^{(k,l)} \equiv {\dis \int^{M_\tau^2}_0}
{\rm d} s \left(1-\frac{s}{M_\tau^2}\right)^k \, 
\left( \frac{s}{M_\tau^2} \right)^l \, \frac{{\rm d} R_\tau}{{\rm d} s}
\ee
can be calculated analogously to $R_\tau=R_\tau^{(0,0)}$
with the QCD OPE. One gets, 
\ba
R_\tau^{(k,l)} \equiv N_c S_{\rm EW} 
\Big\{ (|V_{ud}|^2 + |V_{us}|^2) \,  \left[ 1 + \delta^{(k,l)(0)}\right]
 \nonumber \\
 + {\dis \sum_{D\geq2}} \left[ |V_{ud}|^2 \delta^{(k,l)(D)}_{ud}
+ |V_{us}|^2 \delta^{(k,l)(D)}_{us} \right] \Big\} \, . 
\ea
The electroweak radiative correction 
$S_{\rm EW}=1.0201\pm0.0003$ \cite{Sew} has been pulled out explicitly
and $\delta^{(k,l)(0)}$ denotes  the purely perturbative
dimension-zero contribution.  The symbols $\delta^{(k,l)(D)}_{ij}$
stand for  higher dimensional corrections in the OPE from 
dimension  $D\geq 2$ operators which contain  implicit $1/M_\tau^D$
suppression factors \cite{BNP92,PP98,PP99,CK93}.

The dominant  SU(3) breaking corrections in (\ref{OPE})
are  the dimension $D=2$ quark mass squared terms and
the dimension $D=4$ terms proportional 
to $m_s \langle \overline q q \rangle$.
  The separate measurement of the Cabibbo-allowed
and Cabibbo-suppressed  $\tau$ decay widths 
allows  to quantify this SU(3) breaking through the 
difference
\ba
\label{deltaR}
\delta R^{(k,l)}_\tau \!\!&\equiv\!\!& 
\frac{R^{(k,l)}_{\tau,V+A}}{|V_{ud}|^2}-
\frac{R^{(k,l)}_{\tau,S}}{|V_{us}|^2}  \\ &=& 
N_c \, S_{EW} \, {\dis \sum_{D\geq 2}}
 \left[ \delta^{(k,l)(D)}_{ud}
-\delta^{(k,l)(D)}_{us}\right] \, . \nonumber
\ea
 These observables vanish in the SU(3) limit which, apart from enhancing
 the sensitivity to the strange quark mass, 
also help  to reduce many theoretical uncertainties
which vanish in that limit.

 The dimension two corrections $\delta^{(k,l)(2)}_{ij}$
are known to ${\cal O}(a^3)$ for both $J=L$
and $J=L+T$ components --see \cite{PP98,PP99,CHE04} for references.  
The ${\cal O}(a^3)$ correction
for the $J=L+T$ component has been
presented at this workshop \cite{CHE04}. The results
obtained in \cite{GJPPS04a} and discussed here do not contain these
${\cal O}(a^3)$ corrections and a full
 analysis taking them into account
  will be presented elsewhere \cite{GJPPS04b}.

 An extensive analysis of the perturbative series
for the dimension two corrections was done in
\cite{PP98}. The conclusions there were that  while the 
perturbative series for $J=L+T$ converges very well
the one for the $J=L$ behaves very badly.

In the following applications, the dimension four corrections 
$\delta^{(k,l)(4)}_{ij}$ are fully included while the
dimension six corrections $\delta^{(k,l)(6)}_{ij}$ were
estimated to be of the order or smaller than the error
of the dimension four  \cite{PP99} and therefore will
not be included.

\section{Fixed $m_s$: Determination of $|V_{us}|$}

Taking advantage of the large sensitivity of the
SU(3) quantities $\delta R^{(k,l)}_\tau$ to $|V_{us}|$,
one can obtain a determination of this Cabibbo-Kobayashi-Maskawa
(CKM) matrix element
 using as input a fixed value for $m_s$.
We use as input value $m_s( 2 \,{\rm GeV})=(95\pm20)$ MeV
which includes the most recent determinations of $m_s$
from QCD Sum Rules \cite{MK02,JOP02,NAR99}, 
lattice QCD \cite{HAS04} and $\tau$ hadronic data
 \cite{CHD97,PP98,CKP98,PP99,KM00,CDGHPP01,GJPPS03}.
With this strange quark mass input, one can calculate 
$\delta R^{(k,l)}_\tau$ in (\ref{deltaR}) from theory.

In order to  circumvent the problem of the bad QCD behavior of the 
$J=L$ component in $\delta R^{(k,l)}_\tau$, we replace the QCD
expression for scalar and pseudo-scalar correlators by 
the corresponding phenomenological hadronic parameterizations
\cite{GJPPS03}.   In particular, the 
pseudo-scalar spectral functions are dominated by far
by the well known kaon pole  to which we add suppressed
contributions from the pion pole as well as 
higher excited pseudo-scalar states whose parameters have
been  estimated in \cite{MK02}.

For the strange scalar spectral function  we take
the one obtained from a study of S-wave $K\pi$ scattering
\cite{JOP00} in the framework of  resonance chiral
perturbation theory \cite{EGPR89}
and used in \cite{JOP02} in a scalar QCD sum rule determination
of the strange quark mass.  For comparison, we show in Table
1 the results obtained for the different components
either using the OPE or the phenomenological hadronic parameterizations.
 Being both compatible, the uncertainties are much smaller for the
phenomenological results which we therefore take to replace the
corresponding OPE  contributions to $\delta R^{(k,l)L}_\tau$
while we take $\delta R^{(k,l)L+T}_\tau$  from the QCD OPE
as mentioned above \cite{PP99}.
 As a result the final theoretical uncertainty for
$\delta R^{(k,l)}_\tau$ is much reduced.
\begin{table*}[htb]
\label{table1}
\caption{Comparison between the OPE and the phenomenological
hadronic parameterizations explained in the text
 for the longitudinal component of $R^{(0,0)}_{\tau,V/A}$.}
\begin{center}
\begin{tabular}{cccc}
\hline 
& $R^{(0,0)L}_{us,A}$ & $R^{(0,0)L}_{us,V}$ & 
$R^{(0,0)L}_{ud,A}\times 10^{3}$ \\
\hline 
OPE & $-0.144\pm0.024$ &$-0.028\pm0.021$ & $-7.79\pm0.14$ \\
Pheno. & $-0.135\pm0.003$ &$-0.028\pm0.004$ & $-7.77\pm0.08$ \\
\hline
\end{tabular}
\end{center}
\end{table*}

 The smallest theoretical uncertainty arises for the moment
$(k,l)=(0,0)$, for which we get
\ba
\delta R^{(0,0)}_{\tau,{\rm th}}
&=& (0.162\pm0.013) + (6.1 \pm 0.6) \, m_s^2 
\nonumber \\
&-&(7.8\pm0.8) \, m_s^4 = 0.218 \pm 0.026
\ea
where   $m_s$ denotes the strange quark mass in MeV units,
defined  in the $\overline{MS}$ scheme  at 2 GeV.
 Here, one observes  explicitly the relatively small
sensitivity of $\delta R^{(0,0)}_{\tau,{\rm th}}$
to the actual value of the strange quark mass once
 the longitudinal  component has been substituted by
its phenomenological parameterization.
We have also replaced the
leading dimension four correction $\langle m_s \overline s s 
- m_d \overline d d \rangle$ by its chiral
perturbation theory (CHPT) expression \cite{PP99}.

 OPAL has recently updated the strange spectral function
in \cite{OPAL04}. In particular, they measure
a larger branching fraction $B(\tau^-\to K^- \pi^+ \pi^- \nu)$
which agrees with the previously one measured by CLEO \cite{CLEO03}.   
From the OPAL data and using
\be
|V_{us}|^2 = 
\frac{R^{(0,0)}_{\tau,S}}
{\frac{\dis R^{(0,0)}_{\tau,V+A}}
{\dis |V_{ud}|^2}-\delta R^{(0,0)}_{\tau,{\rm th}}} \, ;
\ee
we get the result
\ba
\label{vus}
|V_{us}|&=&0.2208\pm0.0033_{\rm exp}\pm 0.0009_{\rm th}\nonumber
\\
&=& 0.2208 \pm 0.0034 \, ; 
\ea
where we have used as input strange quark mass 
the one discussed at the beginning of this section and 
the PDG value for $|V_{ud}|=0.9738\pm0.0005$.
Clearly, the uncertainty for the $|V_{us}|$
determination with hadronic $\tau$  
decays becomes an experimental issue
which certainly will be much reduced  at the BABAR and BELLE
B-factories.

One remark is in order here,  the  branching fraction
$B(\tau \to K\nu (\gamma) )$
can be predicted using its relation to the $K_{\mu 2}$ branching fraction.
This relation  is under rather good theoretical control \cite{DF94}.
Updating  the numerics used there, one gets 
$B(\tau \to K\nu (\gamma) )=(0.715\pm 0.004) \%$ which is more precise
than the present world average 
$B(\tau \to K\nu (\gamma) )=(0.686\pm0.023)\%$.
Using this theoretical prediction one gets  $ |V_{us}|=0.2219\pm0.0034$.
 A new more precise determination of this branching
fraction would be very welcomed
and certainly attainable at the current B-factories. 

\section{Fixed $|V_{us}|$: Determination of $m_s$}

One can now use the value obtained for $|V_{us}|$
in the previous section
in a determination of $m_s$ using the  updated OPAL
spectral function. This procedure is meaningful  since 
the moment $(0,0)$ is much more sensitive to 
 $|V_{us}|$ than to $m_s$. We again replace
the phenomenological result for $\delta R^{(k,l), L}_{\tau, {\rm phen}}$
by its OPE counterpart.
Thus we use  \cite{PP99}
\ba
m_s^2(M_\tau)
\simeq \frac{M_\tau^2}{1-\varepsilon_d^2}
\frac{1}{\Delta^{L+T(2)}_{(k,l)}(a_\tau)}\nonumber \\
\times
       \left[ \frac{\delta R^{(k,l)L+T}_\tau} {18 S_{EW}}
+ \frac{8}{3}\pi^2 \frac{\delta O_4(M_\tau)}{M_\tau^4}
Q_{(k,l)}^{L+T} (a_\tau) \right] \, ; 
\ea
with 
$\delta O_4(M_\tau)
 \equiv \langle m_s \overline s s - m_d \overline d d \rangle$,
$\varepsilon_d \equiv m_d/m_s$
 and as input $\delta R^{(k,l)L+T}_\tau= \delta R^{(k,l)}_\tau-
\delta R^{(k,l)L}_{\tau,{\rm phen}}$.

The  term $\Delta^{L+T}_{(k,l)}(a_\tau)$ is
 known in perturbative QCD and the convergence of the
series is very good to ${\cal O }(a^2)$ \cite{PP98}.  
The term  $Q_{(k,l)}^{L+T}(a_\tau)$ is also known
in perturbative QCD to ${\cal O}(a^2)$ --see \cite{PP99}
for references.
The effect of the new ${\cal O}(a^3)$ corrections
to $\Delta^{L+T}_{(k,l)}(a_\tau)$ presented at this
workshop \cite{CHE04} will be investigated in \cite{GJPPS04b}.

 The results we get for the different moments are shown in Table 2
--for the sources of the individual errors see \cite{GJPPS04a}.
The moments $(0,0)$ and $(1,0)$ produce results
with uncertainties larger than 100\% due to their small
sensitivity to $m_s$ and therefore do not contribute to the final
weighted average.
\begin{table*}[htb]
\label{table2}
\caption{Results for $m_s(M_\tau)$ extracted from the different
moments.}
\vspace*{0.3cm}
\hspace*{3cm}
\begin{tabular}{cccc}
\hline 
Moment & (2,0) & (3,0) & (4,0) \\
\hline 
$m_s(M_\tau)$ MeV & $93.2^{+34}_{-44}$ & $86.3^{+25}_{-30}$ 
& $79.2^{+21}_{-23}$ \\
\hline
\end{tabular}
\end{table*}

The weighted average of the strange mass values
obtained for the different moments give
\ba
\label{ms}
m_s(M_\tau)= 84 \pm 23 \, {\rm MeV} \, , \nonumber \\
\Rightarrow  m_s(2\,{\rm GeV})= 81 \pm 22 \, {\rm MeV} \ .
\ea
The final uncertainty corresponds
to that of the $(4,0)$ moment. The dominant  theoretical
uncertainties originate from higher order perturbative
corrections as well as the SU(3)-breaking ratio
of the quark condensates 
$\langle \overline s s \rangle / \langle \overline d d \rangle=
0.8\pm0.2$  \cite{JAM02}.

There are two clear features  
of the result with respect to the analysis
using the ALEPH spectral function \cite{GJPPS03}:
 first, the strong $(k,0)$-moment dependence is reduced
and second, a somewhat reduced but fully compatible 
value for the strange quark mass.
They are both  related to the larger OPAL and CLEO branching fraction 
$B(\tau^-\to K^-\pi^+ \pi^- \nu)$.
This manifests the very important task of the B-factories reducing the
uncertainties in the strange spectral function.

\section{Combined Fit to $|V_{us}|$ and $m_s$}

 The ultimate procedure to determine $|V_{us}|$ and
the strange quark mass from $\tau$ hadronic data will be
a simultaneous fit of both to a set of moments. A detailed
study  including theoretical and experimental correlations
will be presented elsewhere \cite{GJPPS04b}.

 The first step toward this goal was already presented
in \cite{GJPPS04a}. There, we neglected  all correlations
and use the five OPAL moments \cite{OPAL04} from 
$R^{(0,0)}_\tau$ to $R^{(4,0)}_\tau$. The result of this 
exercise is
\be
|V_{us}|=0.2196 \hspace{0.5cm} {\rm and} \hspace*{0.5cm} 
m_s(2 \,{\rm GeV}) = 76 \, {\rm MeV} \, . 
\ee
 These values are in very good agreement  with the  results
(\ref{vus}) and (\ref{ms})
obtained in the previous sections.  The expected final uncertainties
are expected  to be somewhat smaller than the individual errors
for those  but only slightly
since the correlations between different moments are rather 
strong.

\section{Results and Conclusions}

 The high precision hadronic $\tau$ Cabibbo-suppressed 
decay data from ALEPH and OPAL at LEP and CLEO at CESR
provide already competitive results for $|V_{us}|$
and $m_s$. Using the strange spectral function
updated by OPAL \cite{OPAL04}, we get
\be
|V_{us}|=0.2208\pm0.0033_{\rm exp}\pm 0.0009_{\rm th} \, , 
\ee
and 
\be
\label{ms2}
m_s(2 \,{\rm GeV}) = 81\pm 22 \, {\rm MeV} \, ;
\ee
as discussed in Sections 2 and 3 respectively.
The  combined fit to determine both is under way
and will be ready soon. 

 The actual status of the $|V_{us}|$ determinations
 has been nicely reviewed recently  in \cite{CMS04,ROS04}.
 Though the CKM unitarity  discrepancy has certainly
decreased with the new theoretical and experimental
advances, the situation is not yet as good as one could
wish.  

The classical way of determining
$|V_{us}|$ has been through $K_{e3}$ decays.
The uncertainty of this determination
 is dominated by the theoretical prediction of the form factor $f_+(0)$,
which is known model independently just to ${\cal O}(p^4)$
in CHPT.  Recently, there have appeared several works 
updating the old model calculation of the ${\cal O}(p^6)$
corrections  to $f_+(0)$ by Leutwyler and Roos \cite{LR84}.
 The present status of this form factor can be 
inferred from the calculation in \cite{JOP04} using 
resonance chiral perturbation theory and from
the calculation in \cite{BILMMSTV04}
which presented a quenched lattice calculation 
of the  order $p^6$ corrections.  
 $|V_{us}|$ can be also determined
from hyperon decays \cite{CSW03,FLO04}.
Another recent proposal has been to
use the lattice QCD determination of $f_K/f_\pi$
\cite{MAR04,MILC04}. 
The accuracy of all these determinations is  still in the
$\geq$ 1\% range for the uncertainty, which will be 
 difficult to decrease at short or even medium term. 
One long term possibility to reduce the
error in predicting $f_+(0)$ is the proposal in \cite{BT03}.

Thus, there is room for hadronic $\tau$ decays to make an
accurate  determination for $|V_{us}|$
 with the eventual accurate measurement of
the strange spectral function at BABAR and BELLE.

 Our result for the strange  quark mass is on the low
side of previous determinations, but certainly compatible
with them. It is also borderline of being compatible
with lower bounds on $m_s$ from sum rules 
\cite{MK02,LRT97,YND98,DN98,LS98}. It is also compatible
with the sum rules determinations of $2 \hat m \equiv m_u+m_d$ 
which can be  combined with the ratio $m_s/ \hat m$ from
CHPT determinations. If one uses for instance
the result for $m_u+m_s$ from \cite{BPR95} and the ratio from
\cite{LEU96} one gets $m_s(2 \,{\rm GeV})=114\pm22$ MeV which
is just  one $\sigma$ from (\ref{ms2}).

 There are however several open questions that will have also
to be addressed. The  $(k,0)$-moment dependence of the 
$m_s$ prediction has been reduced after the recent
OPAL and CLEO analyses finding larger branching fractions
for $\tau^- \to K^- \pi^+ \pi^- \nu$. That this  moment
dependence could be due 
 to missing contributions from the higher energy region
of the spectrum was speculated in \cite{JAM03}. What is the
origin of the remaining moment dependence is still open.

In \cite{CDGHPP01}, it was checked that 
 the $m_s$ determination fulfills quark-hadron duality 
between the QCD OPE and the ALEPH data. What happens with $|V_{us}|$
is another open question. A first look at this question has been
presented at this workshop \cite{MAL04}. 
We will eventually come back to all these questions
in \cite{GJPPS04b}. 

\section*{Acknowledgments}
 This work has been supported in  part by
the European Union (EU) RTN Network EURIDICE
Grant No. HPRN-CT2002-00311 (A.P. and J.P.),
by German BMBF under Contract No. 05HT4WOA/3 (M.J.),
by MCYT (Spain) and FEDER (EU) Grants No.
FPA2001-3031 (M.J. and A.P.) and FPA2003-09298-C02-01 (J.P.),
  and by Junta de Andaluc\'{\i}a
Grant No. FQM-101 (J.P.). E.G. is indebted
to the EU for a Marie-Curie Intra-European Fellowship.

\end{document}